# Oblivion of Online Reputation: How Time Cues Improve Online Recruitment

Alexander Novotny* and Sarah Spiekermann

Institute for Management Information Systems,
Vienna University of Economics and Business Administration,
Welthandelsplatz 1
A-1020 Vienna
Tel: +43 1 313 36 4428
Fax: +43 1 313 36 904428
E-Mail: alexander.novotny@wu.ac.at
E-Mail: sarah.spiekermann@wu.ac.at
*Corresponding author

**Abstract:** In online crowdsourcing labour markets, employers decide which job-seekers to hire based on their reputation profiles. If reputation systems neglect the aspect of time when displaying reputation profiles, though, employers risk taking false decisions, deeming an obsolete reputation to be still relevant. As a consequence, job-seekers might be unwarrantedly deprived of getting hired for new jobs and can be harmed in their professional careers in the long-run. This paper argues that exposing employers to the temporal context of job-seekers' reputation leads to better hiring decisions. Visible temporal context in reputation systems helps employers to ignore a job-seeker's obsolete reputation. An experimental lab study with 335 students shows that current reputation systems fall short of making them aware of obsolete reputation. In contrast, graphical time cues improve the social efficiency of hiring decisions.

**Keywords:** obsolete reputation, online crowdsourcing labour markets, right to be forgotten

**Biographical notes:** Alexander Novotny is an information privacy and security specialist. He works as an information security risk manager in the utilities industry. Moreover, he is a lecturer at the Institute for Management Information Systems at the Vienna University of Economics and Business. He has been teaching lectures on information privacy and security, ethical computing, and the foundations of information and communication technology. Alexander holds a Ph.D. in the economic and social sciences with a major in business information systems. His research interests focus on electronic privacy, information security and ethical computing. He served as a standardization expert for digital marketing and privacy at the Austrian Standards Institute.

Sarah Spiekermann is a professor for Information Systems and chairs the Institute for Management Information Systems at Vienna University of Economics and Business. Before tenured in Vienna, she was assistant professor at the Institute of Information Systems at Humboldt University Berlin, Germany and held an Adjunct Professor positions with Carnegie Mellon University. Sarah is best known for her work on electronic privacy and electronic marketing. The key goal of her work is to investigate the importance of behavioral constructs and social values for IT design and to refine the concept of ethical computing in an e-Society. She co-founded the Privacy & Sustainable Computing Lab at the Vienna University of Economics and Business.

## 1 Introduction

A good online reputation is increasingly vital for people. Online reputation systems retain the performance histories of people for an indefinite time and determine to a large extent "what is generally said or believed about a person's or thing's character or standing" (Jøsang et al., 2007, p.620). Take the case of Med Express vs. Amy Nichols (Wolford, 2013). Following a dispute on shipping costs, Mrs. Nichols left a negative review on eBay seller Med Express. Since Med Express was aware that the negative feedback would persistently stay in its reputation profile, it sued Nichols for having caused irreparable damage of lost customer value and revenue. Indeed, market actors with a low reputation receive substantially lower prices (Depken Ii and Gregorius, 2010). Eventually, Med Express dismissed the lawsuit.

While online reputational feedback is indefinitely retained online, people's true being and careers are marked by progress, disruptions and shifts. On average, U.S. citizens have 11 different jobs throughout their working lives (BLS, 2012) and move their residences 12 times during a lifetime (USCensus, 2007). With bankruptcy filers increasing from 0.15 to 5.3 per 1,000 people and year over the last century, people's financial lives are increasingly coined by crisis and restart (Garrett, 2006). Constant human change is not only limited to such externally observable professional life changes. Aristotle argues in his "Physics" that no time can pass without humans' inner change (Coope, 2001). Ricoeur's (1988) phenomenological philosophy comes to a similar conclusion. In his narrative model, he argues that novel experiences as well as new connections and relationships between actors surfacing over time lead people to a constant reframing of their histories and pave the way for personal evolvement. This constant human evolution stands in sharp contrast to the static and almost timeless online presence of people. A problem arises when statically retained online information diminishes a person's true reputation and arbitrarily harms him or her.

To protect people from such a damaged reputation, EU policy makers decided to include a "right to be forgotten" in Europe's Data Protection Regulation (De Hert and Papakonstantinou, 2012, p.136). The right aims at enabling a person to "determine the development of his life in an autonomous way, without being perpetually or periodically stigmatized as a consequence of a specific action performed in the past" (Mantelero, 2013, p.230). The legal metaphor of "forgetting" on the Internet refers to how prominently and easily a person's online information can be accessed. Of course, society should never forget some of people's historic actions, for example the atrocities by some dictators and their henchmen. But apart from such consciously unforgotten crimes, personal information retained online may be forgotten after some time.

In this article we investigate the problem of obsolete online reputation for online recruiting, in particular in scope of online crowdsourcing labour markets. On globally operating crowdsourcing labour market platforms, job-seekers are domiciled in different countries. Particularly through the job-seeker's reputation profiles on these platforms, employers become acquainted with job-seekers and take a decision whether to hire them. Examples are platforms such as upwork.com, elance.com and guru.com. They use an "open call format" to request work from a "large network of potential labourers" (Howe and Robinson, 2006). Typically, the work requested is on a project basis. They may involve the creation of software, graphic works and advertising texts, for instance.

One sort of software works sourced from online crowdsourcing labour markets are app designers developing mobile apps customized to the needs of their employing principals. Apps are software programs that are particularly designed to run on mobile devices such as smartphones and tablet PCs. The demand for developing apps facing customers or for enterprise purposes is rising (GoodTechnology, 2014). Customer apps, for example, allow browsing product catalogs and finding the next store while on the go. Enterprise apps may support travelling employees with access to documents and the company's customer database. Compared to traditional offline recruitment channels, online crowdsourcing labour markets offering job-seeking app designers recruitable for app development projects typically address employers who are more technology-affine. One could argue that online crowdsourcing labour markets in particular should strive for providing employers with the timeliest information about potential job-seekers. Temporal context information about how a job seeker is today versus how he or she was in the past is crucial. Avoiding that employers hire based on obsolete reputation thereby harming job-seekers and society, studies advocate for employers disregarding obsolete reputation: "since the reputation values are associated with human individuals and humans change their behaviour over time, it is desirable to disregard very old ratings" (Zacharia et al., 2000, p.376). Also, older online reviews are perceived to be less helpful for making decisions on online market transactions (Cao et al., 2011; Hu et al., 2008).

However, today's online crowdsourcing labour markets make the importance of time little visually salient yet. Employers are easily led to neglect the necessity to disregard obsolete reputation when they judge on potential new hires.

The reason for this neglect is that current user interfaces of crowdsourcing labour markets' reputation systems do not make employers aware of reviews that are outdated. In previous work (source blinded for peer-review), 16 generic categories of design alternatives were found that help visualizing time context of person-related information in user interfaces. Design alternatives for time visualization, for example, include graphical timelines and visual decay. Timelines display the time-related dimension of information on horizontal screen-space. Interfaces drawing on the visual decay metaphor present outdated information in a progressively dissolved state. Current user interfaces of crowdsourcing labour markets' reputation systems, though, only use graphical time visualizations assigned to one category: text-based symbol of time. These interfaces, for instance, add a timestamp to reviews and display text-based listings of the feedback received within the last 12 months. They only contain tiny text-based cues to the publication date, mostly printed in small font size and in grey letters. None of them, though, uses any graphical time visualizations (see Table 6 in the Appendix). HCI studies have shown for long that text cues have little salience in interfaces and impose an additional cognitive load on profile viewers (Hong et al., 2007). As a result, employers can probably hardly notice the temporal context of reviews.

In addition to textual time cues, current reputation systems optionally sort reviews by publication time (see Table 6 in the Appendix). Empirical results show that chronologically ordering reviews increases their helpfulness (Otterbacher, 2009). Like on a job-seeker's curriculum vitae (CV), recent occurrences are displayed first, making them immediately visible to employers. eBay's reputation system, for example, provides sophisticated options for sorting and aggregating feedback by time. Feedback can be filtered by recentness (last 1, 6 and 12 months) keeping all feedback still accessible. But recency filters can be misleading as well. If someone has mostly performed well in the past, but recently failed to meet some demands, chances are high that job seekers immediately suffer penalties, because their history is not viewed holistically.

Against this background, we argue in this paper that reputation systems should more saliently visualize time. Visibility is generally a key design principle of user interfaces aiming to keep information recipients on top of the current status of a system, process, action or entity (Nielsen, 1994, p.154; Norman, 1988). We argue that a higher visibility of reviews' temporal aspects makes future employers aware of outdated reputational information and helps them to avoid it. Potential future employers are less likely to base their judgments on obsolete reputation. Aiming to address the call for a "right to be forgotten" from a system design angle, we ask: How should reputation systems display obsolete information? Would a higher visibility of reputation's temporal context encourage employers to avoid obsolete reputation and focus on job-seekers' timely achievements? And would first-rate job-seekers who are still plagued by a negative, outdated reputation obtain a second chance of getting hired?

To elucidate these questions, an experimental laboratory study was conducted. It manipulated visual time cues in the reputation system user interface of a fictitious online crowdsourcing labour market. The experiment manipulated two types of visual time cues: First (and similar to how current online crowdsourcing labour markets' reputation systems organize time) the temporal order of reviews in reputation profiles was manipulated. Second, we tested the effects of a new salient graphical timelines cue in reputation profiles. We find that graphical timelines encourage employers to assess job-seekers in a more recent light and to discount job-seekers' obsolete reputation. In contrast, we could not find this effect for ordering reviews by time which is the current standard for job-seekers' reputation profiles.

The remainder of this paper is structured as follows. Section 2 presents the theoretical background and hypotheses. Section 3 describes the method of the experimental study. Section 4 outlines the results which are discussed in Section 5. The final Section 6 draws conclusions and points to future work.

**2 Theoretical background**

In online crowdsourcing labour markets, employers face the problem of hiring job-seekers without having knowledge of their skills, traits, experience and quality of work. Information asymmetry (Akerlof, 1970) reigns between employers and job-seekers making hiring decisions risky. This section reviews how online reputation systems help reducing this information asymmetry and handle temporal aspects of job-seekers' reputation. Then, we outline our hypotheses how time cues for reputation can support employers with disregarding obsolete reputation.

*Online reputation systems and market transparency*

To reduce uncertainty about job-seekers' quality, online crowdsourcing labour markets operate reputation systems (Jøsang et al., 2007). "A reputation system collects, distributes, and aggregates feedback about [market] participants' past behaviour" (Resnick et al., 2000). A good online reputation serves as a trust mark reducing transaction risks and remediating information asymmetry (Ba and Pavlou, 2002). The trust-building effect and transparency for employers is highest if they have access to the full reputation history of job-seekers. Maximum market transparency is achieved if no incidents in a job-seeker's reputation history are omitted. Hence, deleting older reviews in job-seekers' reputation profiles would reduce transparency.

Also, allowing job-seekers to delete reviews could incentivize them to manipulate their reputation and cast themselves in a potentially falsified positive light. Such feedback manipulation undermines the very goal of online reputation systems, that is to present a person's true background (Dini and Spagnolo, 2009). It would also censor past employers' opinions and their right to free speech online (Rosen, 2012).

Yet, one dimension of transparency is also to consider the "appropriateness" of information (Spiekermann, 2015; Turilli and Floridi, 2009). It may be that reviews about job-seekers were entered into a reputation system long ago and are by now obsolete. The person concerned may have changed. If deleting such obsolete reviews is no viable option for the reasons given above, how can job-seekers safeguard themselves from being eternally reproached for their past?

For employers appropriately basing their hiring decision on information about the up-to-date current skills and experience of job-seekers, they should disregard job seekers' obsolete past behaviour and "forget" it at some point in time. Zacharia et al. (2000, p.376) argue: "Reputation values are associated with human individuals and humans change their behaviour over time, it is desirable to disregard very old ratings." At the organizational level, employers are therefore well advised to disregard those parts of a job-seeker's reputation which seems to be obsolete.

Ignoring the need to disregard obsolete reputation bears negative consequences at the societal level. Employers can run the risk of seeing job-seekers in a "false light" (Prosser, 1960, p.398). They assess job-seekers based on outdated reputational information that does not accurately reflect the job-seeker's current skills and quality. If reputation in electronic markets is perceived to be low, lower market prices (i.e., wages) can result (Depken Ii and Gregorius, 2010). In extreme cases, actors with a low, but obsolete online reputation may get "stigmatized" (Solove, 2006, p.547). Such job-seekers may never get hired in the online crowdsourcing labour market again. Rebuilding a positive reputation for the future is tedious and costly (Matzat and Snijders, 2012). Eventually, despite their good present work quality that would rehabilitate an obsolete negative reputation, they may have to drop out of the job market. Consequently, not disregarding obsolete reputation may create social costs in the crowdsourcing labour market.

If we embrace the argument that obsolete reputation should be disregarded and "forgotten" online, then it would make sense to support employers with this task. Reputation systems should then facilitate and remind employers to disregard outdated reputation.

User interface cues may help employers to identify and ignore obsolete reputation. User interface cues were found to influence the behaviour of actors in electronic market environments: for example, visually cueing to the human face of sales personnel was found to increase consumer trust on e-commerce websites (Aldiri et al., 2008). Also, the design of reputation systems was shown to influence the effect of a market actor's reputation on a transaction (Klein et al., 2009; Zhu and Zhang, 2010). In online political discussion fora, visual interface cues were found to make online community members listen better to each other's opinions (Manosevitch et al., 2014). The design of online reputation systems may also cue reviewers to rethink their behaviour (Ekstrom et al., 2005).

*Obsolescence of reputation*

If user interfaces should cue employers to obsolete reputation, when is a job-seeker's online reputation *obsolete* in the particular context under investigation? In an online survey about reputation in crowdsourcing labour markets (n=494, the participants were sampled from the same student population as in the experiment), we found that 82.8% of reputation system users believe that reviews about job-seekers are outdated if these were published three years or longer ago (see Figure 1). Individual employers gradually deem reviews to get obsolete based on various factors such as job seekers' ability to learn, gain experience and their changing motivation and personality. Even though individual perceptions of review obsolescence vary depending on these factors, there is a community consensus on when reviews shall be considered as outdated and not be further used to judge job-seekers. In our analysis presented below we make the conservative assumption that reviews published more than four years ago are obsolete (even 93.3% of reputation system users deem five year old reviews to be obsolete). Hence, employers should mainly consider reviews which were published less than four years ago to get a rather recent and accurate picture of job-seekers. To support employers with this task, reputation systems should highlight the temporal context of reputation.

*Hypotheses*

Despite the importance of disregarding obsolete reputation for hiring, some current online crowdsourcing labour market platforms tend to prefer highlighting the best achievements of app designers over their most recent ones (see Table 6). Consequently, these platforms display reviews by ordering the highest rated ones first in reputation profiles. Employers are advertised that chances of finding a well-rated job-seekers are high on the respective platform. In line with our argumentation, though, we expect that this practice disobeys time aspects and does not encourage employers to disregard obsolete reputation.

*H1: Reputation profile interfaces ordering the reviews by star rating do not influence employer disregard of obsolete reputation.*

On the contrary, this paper proposes that visually cueing to reviews' temporal context in a reputation system's user interface (time cues) hints employers to disregard obsolete reputation. Because of their higher awareness of the reviews' temporal context, they can become more aware that obsolete reviews are irrelevant and inappropriate for their hiring decisions.

We hypothesize that two types of visual time cues that reputation systems may include increase employers' disregard of obsolete reputation: temporal order and graphical timelines. The temporal order cue represents the current practice of how online crowdsourcing labour markets' reputation systems focus employers on time (see Table 6 in the Appendix). Following this current practice, we expect temporal order cues encourage employers to disregard obsolete reputation when hiring:

*H2: Reputation profile interfaces ordering the reviews by time increase employer disregard of obsolete reputation.*

Chronologic CVs used by job seekers to traditionally apply for jobs follow the convention of being in descending temporal order. Employers start reading CVs from the top, thereby first encountering a job seeker's most recent experiences. Also, narrating job-seekers' reputation histories in reputation profiles can either start with the most recent or the oldest event. We expect that online crowdsourcing labour market platforms sorting the most recent reviews first in reputation profiles guide employers' attention closer to job-seekers' recent professional conduct. As a result, employers are more likely to disregard obsolete reputation when making hiring decisions.

*H3. Reviews in descending temporal order more strongly encourage employers to disregard obsolete reputation than reviews in ascending temporal order.*

Beyond mere time order, this paper suggests a novel time cue for reputation systems: timelines that graphically visualize the passage of time in reputation profiles. Online advertising effectiveness research shows that graphical interface cues attract more attention than text-based or mixed cues (Hsieh and Chen, 2011). Visualizing time is a largely unexplored topic in human-computer interaction (HCI) research (Lindley et al., 2013, p.3212). Keeping with how Western cultures read from left to right, proposals use timelines mapping the past to the left and the future to the right (Santiago et al., 2007). Timelines are visual metaphors capable of narrating personal histories (Plaisant et al., 1996; Thiry et al., 2013) and could also be used to tell job-seekers "reputation histories". The timelines cue was selected over other approaches for graphically visualizing time ([source blinded for peer-review], p. 552) for three reasons. First, timelines are suitable for representing a reviews' publication date with high granularity. Second, they are able to clearly arrange high amounts of reviews contained in reputation profiles. And they can be dynamically created within the user interface ([source blinded for peer-review], p. 553).

The idea of graphical timelines is taken up to visualize when reviews were published within reputation profiles. We introduce a timelines cue representing reviews as dots on graphical timelines progressing from left to right (see Figure 2). The dots contain the review's rating. Timelines aim focusing employers towards recent reputational information about job-seekers' when assessing job-seekers and pointing employers towards disregarding obsolete reputation:

*H4: Reputation profile interfaces displaying reviews on graphical timelines increase employer disregard of obsolete reputation.*

**3 Method**

A computer lab experiment was conducted in German with students who were all online reputation system users. Participants were recruited via a university mailing list. Those who are not reputation system users and who already participated in the preceding survey were denied registration. Participants were paid 10 euros in cash. They completed the experiment on identical screens which were sheltered from other participants' gazes. Median completion time was 20.92 minutes.

Participants were situated into an identical scenario. They were told that their university wants to develop a new mobile app called the "CampusApp". This app would enable spontaneous orientation on campus and provide ad-hoc directions to currently vacant student project rooms. The interface of a fictitious online crowdsourcing labour market platform showed the participants the reputation profiles of four app designers (hereafter denoted AD1 to AD4) who were offering their support to develop this app. The participants were requested to suggest hiring one out of the four app designers. We asked them: "Who would be the best app designer for the CampusApp?"

A choice set of four profiles reflected hiring decisions between multiple job-seekers on real online crowdsourcing labour market platforms while keeping the selection task manageable for participants. Participants were told that they are in charge of hiring an app designer for the university's app design project. Ensuring incentive compatibility, participants could win additional 20 euros for making a selection that fits the university's app design project well. Participants had free choice and were not restricted to follow their own preferences and judgements while making a decision.

*Online reputation profiles*

The reputation profiles had no systematic differences except for the average rating score of the reviews which were regarded to be still relevant and not obsolete (see Table 1). Each profile contained eight reviews that were published in the 5-year time frame between 2010 and 2014. Online reviews for the profiles were selected from 120 review texts. The reviews were written in a professional, evaluative and non-technical language. The reviews' length was 25 to 35 words. Based on Goldberg's (1990, p.1224) personality trait III+ scheme all reviews made statements about an app designer's conscientiousness and motivation while doing design work. For instance, app designers were described to be dependable, punctual, disorganized, or aimless. The reviews were pre-tested for valence (on a scale from 1.00 to 5.00 stars with 5.00 being the best rating), text quality (readability, helpfulness, understandability) and congruence between headline and text. The pre-test data was based on 3,123 rating points provided by 494 participants (see Section 2). Reputation profiles were compiled to exhibit no statistical differences (n=812) of perceived valence ($F=0.459$, $df=3$, $p=0.711$), text quality ($F=0.216$, $df=3$, $p=0.885$) and congruence between headlines and text ($F=2.115$, $df=3$, $p=0.097$).

In order to avoid bias from identifying profile characteristics, randomly generated pseudonyms were used as app designers' names (e.g., "Wtmfc") and former employers (e.g., "uoweoi898"). Profiles had identical dummy profile pictures showing a person's grey silhouette. The grey silhouettes ensured that choice was not biased by the sympathy and appearance of the app designers. An example profile is depicted in Figure 3.

The app designers' total profile rating scores randomly varied within the range of 4.01 to 4.19 stars. Random variation of total rating scores made the hiring task more realistic because job-seekers do not have equal reputation scores in real-world online crowdsourcing labour markets either. Randomization was stochastically independent of the random assignment of participants to interface manipulations. Multinomial logistic regressions confirmed that the random total rating scores did not systematically influence app designer choice.

*Dependent measure*

We operationalise *employer disregard of obsolete reviews* as follows: reviews published in 2010 were deemed to be outdated an obsolete by employers because they are more than four years old (see above). In contrast, the reviews published between 2011 and 2014 are still regarded to be relevant. Thus, we only use the reviews published between 2011 and 2014 to calculate an average rating score of the non-obsolete reviews for each reputation profile. To calculate the score of a profile, the ratings are added and divided by the number of reviews as depicted in Figure 4 (e.g., for AD3: (4.99+4.86+4.81+4.94+1.87+4.92)/6=4.398). This score is the only systematic difference between the four app designers' reputation profiles (see Table 1). The employers' disregard of the obsolete reviews is then measured by the score of the app designer's profile that was selected by the participant. Employers who select a reputation profile assigned a higher score disregard obsolete reputation to a higher degree.

*Independent variables: manipulations of interface time cues*

In a between-subjects design, the user interface displaying the reputation profiles was manipulated to include time cues. Avoiding order bias, the reputation profiles were displayed in random order to participants. Participants were randomly assigned to one of six conditions. All conditions displayed text dates when the reviews were published. Beyond text dates, three treatment conditions (TC1, TC2, TC3) were using visual time cues. TC1 gives a temporal order cue. The most recent reviews in the profiles are ordered first. TC2 combines the temporal order cue with the timelines cue (see Figure 3). Recent reviews are ordered first and graphical timelines of the profiles are displayed. The timelines cue's colour combination (dark yellow/ultramarine blue) had a neutral colour weight on the Kobayashi colour image scale. The timelines cue is always combined with the temporal order cue because items on a timeline are in a natural temporal order. TC3 is identical to TC2, except that the oldest reviews are ordered first. This condition was added to investigate the influence of a particular direction of temporal order (ascending or descending). Three control conditions did not display visual time cues (CC1, CC2, CC3). The control conditions provide no visual time cues. CC1 sorted the reviews randomly. CC2 and CC3 order the highest rated reviews first. CC3 was added to control for bias resulting from the mere presence of graphical cues in the user interface. It orders the highest rated reviews first and displays graphical profile representations that do not contain any temporal information (see Figure 5). The visual interface cues were pretested.

*Post-questionnaire*

After participants saw the manipulated reputation system interfaces and made an app designer choice, they answered a questionnaire. First, they judged the correctness of six statements on time-related information testing whether visual time cues actually made them focus on time (manipulation checks). As a second check, participants were asked whether they focused on the reputation development history of the individual app designers while making their choice. Answers were provided on a five-point Likert scale ("did incorporate strongly" to "did not incorporate at all").

Then, participants answered control variables and demographics. They rated how often they normally read and author reviews online ("never" to "very frequent"). Two types of participant involvement were controlled for. Involvement into the CampusApp scenario (scenario involvement) was measured by five 7-point semantic differentials adopted from the personal involvement inventory (Zaichkowsky, 1985, p.350). Involvement into the task of selecting an app designer (choice involvement) was measured by six items adapted from the new involvement profile (Jain and Srinivasan, 1990, p.597). Some participants may condone mistakes that app designers made in the past. Others may be less lenient and judge app designers who made mistakes as enduringly negative. Hence, participants' forgiveness – whether they have a generally merciful personality – was measured by five items adapted from the "Heartland forgiveness of others" subscale (Thompson et al., 2005, p.358). Participants' rational thinking style while cognitively processing the reputation profiles was controlled for using three items from the rational situation-specific thinking style scale (Novak and Hoffman, 2009, p.60). The control variables exhibit satisfactory psychometric characteristics (see Table 7 in the Appendix). Other control variables did not explain a relevant amount of covariance and were not included into the analysis.

## 4 Results

382 participants completed the experiment. 47 participants were excluded from the sample: One was no reputation system user. Eight were careless responders finishing the study in less than the median time minus 1.3 times the interquartile range. 38 participants wrongfully answered a check on review order in the reputation profiles. Eventually, a sample of 335 participants remained.

*Descriptive results*

Laboratory participants were pre-screened to be online users with experience of reputation systems. They were young (mean age 23.31 years), well-educated (97.6% university students), familiar with online marketplaces (87.4% transact at least once a year online), and mobile phone apps (88.9% use apps at least often, see Table 2). These characteristics are typical for users affine to online recruitingand were reflected in the design of the experimental scenario. The raw data (n*=382) and cleaned sample (n=335) show akin demographic characteristics (see Table 2). The further analysis is conducted on the cleaned sample.

The manipulation checks confirmed that participants in the TCs had a higher awareness of temporal information contained in the profiles than those in the CCs ($\Delta M_{min}$=0.948, SE=0.351, p=0.007 (CC3-TC1); $\Delta M_{max}$=2.536, SE=0.357, p<0.001 (CC2-TC2)), except for the contrast between TC1 and CC1 which closely missed significance ($\Delta M$=0.724, SE=0.380, p=0.059). This difference is significant for participants in the TCs who put more focus on the reputation development history of the app designers than those in the CCs ($\Delta M_{min}$=0.690, SE=0.263, p=0.009 (CC1-TC1); $\Delta M_{max}$=2.059, SE=0.249, p<0.001 (CC2-TC2)). The most effective manipulations are TC2 followed by TC3 which both combine temporal order and timelines cues. Further, a linear regression confirmed that the manipulations were not confounded with an increase in participants' confidence of their app designer choice (Timelines cue: B=-0.071, SE=0.059, p=0.232; Temporal order cue: B=-0.025, SE=0.060, p=0.671).

Employer disregard of obsolete reputation was consistently higher in the TCs than in the CCs. In the CCs, it similarly ranged between 4.392 (SD=0.335) and 4.427 (SD=0.377). The temporal order cue (TC1) increased disregard of obsolete reputation to 4.534 (SD=0.400). Combining temporal order and timelines cues (TC2) caused the highest disregard of obsolete reputation (M=4.760, SD=0.237). Reversing the direction of temporal order (TC3) resulted in a slightly lower disregard of obsolete reputation (M=4.660, SD=0.310). Table 3 summarizes the number of subjects as well as the means and standard deviations of their disregard of obsolete reputation across the conditions. Both the cleaned sample and raw data (asterisked values in Table 3) exhibit stable means and standard deviations of disregard of obsolete reputation.

Because group sizes were unbalanced (see Table 3), linear regression models serving as unbalanced groups ANOVA were used to examine the differences between the conditions. Main effects were effect coded and simple effects (i.e., contrasts between the conditions) were dummy coded.

*The effect of ordering the reviews by star rating on disregarding obsolete reputation*

The mere presence of a graphical element in the interface (comparing CC2 to CC3) did not influence employers' disregard of obsolete reputation ($\Delta M$=0.035, SE=0.064, p=0.583). In a step-wise linear regression accounting for the control variables, sorting by highest rated reviews first in reputation profiles did not increase disregard of obsolete reputation compared to a random review order (B=0.006, SE=0.057, $\Delta F$=0.012, p=0.912).

There were no differences in employers disregarding obsolete reputation between the three CCs, neither on an overall level (F=0.155, df=2, p=0.856), nor when each CC was contrasted with each other (CC1-CC2: $\Delta M$=-0.012, SE=0.066, p=0.852; CC1-CC3: $\Delta M$=0.023, SE=0.070, p=0.747; CC2-CC3: $\Delta M$=0.035, SE=0.064, p=0.583). Hence, to simplify the further analysis, all CC were combined into one control pool (CP).

*The effect of time cues on disregarding obsolete reputation*

Using a step-wise regression procedure, the main effects of the timelines and temporal order cues were tested (see Table 4). Residuals were normally distributed and error variances were homoscedastic. Multicollinearity of the predictors was low ($VIF_{max}$=2.853; variance inflation factors should be below 10). The timelines cue increased disregard of obsolete reputation and explained the biggest share of variance (see step 5.a, B=0.113, SE=0.033, $\Delta F$=36.800, p=0.001). Comparing TC1 to TC2, the effect of the timelines cue is robust if the temporal order cue is held constant (B=0.201, SE=0.061, $\Delta F$=10.681, p=0.001).

In contrast, the main effect of temporally ordering by recent reviews was not significant (B=0.010, SE=0.033, ΔF=0.089, p=0.766). Also when the particular direction of temporal order is ignored, the temporal order cue is not significant (see step 5.b, B=-0.027, SE=0.048, ΔF=0.307, p=0.580). Because timelines naturally order reviews by time, the "timelines" and "temporal order" predictors are moderately multicollinear (VIF of timelines = 7.650, results robust if VIF<10). Moreover, even though significance is only marginally missed, there was no difference in disregard of obsolete reputation between reviews sorted in descending temporal order (TC2) and in ascending temporal order (TC3) (B=0.094, SE=0.049, ΔF=3.678, p=0.058).

Employers forgiving workers a negative obsolete reputation were more disregarding obsolete reputation (B=0.062, SE=0.019, p=0.001). However, neither the timelines cue (B=0.079, SE=0.160, p=0.620) nor the temporal order cue (B=0.036, SE=0.143, p=0.804) influenced employers' forgiveness. Also, participants who addressed the profiles with a rational thinking style were more disregarding of obsolete reputation (B=0.150, SE=0.030, p<0.001). By systematically comparing the profiles they were recognizing that obsolete reputation should be ignored.

Online crowdsourcing labour market platforms ordering reviews by star rating do not contribute to hint employers to disregard obsolete reputation when making hiring decisions (H1 supported, see Table 5). However, the results provide evidence that visual time cues may increase disregard of obsolete reputation. The effect depends on the types of visual time cues provided in reputation system interfaces. Temporal order cues do not significantly hint employers towards the need to disregard obsolete reputation (H2 rejected), regardless whether reviews are ordered in ascending or descending order (H3 rejected). The timelines cue, in contrast, makes employers aware of disregarding obsolete reputation when hiring (H4 supported).

## 5 Discussion and limitations

The findings show that current online reputation systems tolerate hiring in online crowdsourcing labour markets which are based on obsolete reputation. Textual time cues (i.e., reviews publication dates) and sorting reviews by star rating downplay temporal aspects in job-seeker's reputation profiles. They hardly raise employers' awareness for forgotten need to disregard obsolete reputation.

Designers of online crowdsourcing labour market platforms should therefore follow a few guidelines for displaying job-seekers' reputation histories: Most importantly, the reputation systems on these platforms should provide graphical time cues. The timeline metaphor turned out to be useful for graphically visualizing time. It evokes the basic mental picture of how people imagine time in Western societies. Temporal order cues did not influence employers to disregard obsolete reputation, irrespective whether new or obsolete reviews are put to the front. Sorting by best reviews has similarly no effect on disregard of obsolete reputation than a completely random review order. Hence, reputation systems should avoid this order to be the default option.

Interface designers using visual time cues, though, need to pay attention to usability. Despite no effect of temporal order cues was found, the possibility to sort reviews by time is a feature that reputation system users expect from familiar reputation systems. This feature may be provided additionally in user interfaces.

If user interfaces were designed by these guidelines for presenting obsolete reputation, we argue that job-seekers would regain new chances for job opportunities, despite their profiles containing a negative, obsolete reputation. As a result, online crowdsourcing labour markets would grant job-seekers a fair equality of employment opportunities over the course of their reputational online life. An obsolete reputation would not further irreversibly "knock out" job-seekers from online crowdsourcing labour markets. Job-seekers would have no need to find strategies for declaring "reputation bankruptcy" (Zittrain, 2010, p.228) harming market transparency. Honest job-seekers would not be pressurised migrating to other platforms where their reputation is unknown. And crooked job-seekers would have a lower incentive to delete their accounts containing their reputation profile and re-registering using a false identity. Desires for deleting entire reputation histories would diminish. For ensuring an effective reputation system, only defamatory and denigratory reviews would require deletion.

The visibility created around time-related aspects of reputation also makes job-seekers' recent performance more transparent. Job-seekers whose performance dropped recently may oppose this new transparency because they would need to take over accountability for their recent performance. Our position is that online crowdsourcing labour markets should not support job-seekers with hiding low professional performance that still has relevance for employers in the presence. It is important that job-seekers are not eternally reproached for an obsolete performance that is not further relevant. If employers overestimate a negative, but obsolete reputation of job-seekers, then these job-seekers will not get hired despite their good present work quality. Job-seekers are hindered from recovering from obsolete reputation and eventually are lost for the market. As a result, online crowdsourcing labour markets forfeit a share of their efficiency.

An unanticipated result is that employers' forgiveness plays a key role when job-seekers' reputation is retained forever. Forgiving and merciful employers were more disregarding obsolete reputation contained in the profiles. Forgiveness, though, is a predisposition of employers and was not influenced by the interface cues. This finding can be interpreted as more merciful employers ignoring differences in job-seekers' individual reputation histories after some time. For merciful employers, a job-seeker's obsolete reputation is no longer relevant.

*Limitations*

Our study has several limitations. The university student sample of reputation system users enables drawing conclusions about a young, technology-affine population who is familiar with online channels for reputation. It does not, though, allow generalizing the results to other settings, particularly to offline contexts. The visual time cues were tested using a Western culture sample. Reputation system users from other cultural areas may have a different mental picture of time. Moreover, our results on the time period until obsolescence of online reputation is highly context-dependent within the setting of reputation systems on online crowdsourcing labour market platforms. In other contexts, such as on e-shopping or online auction platforms, the appropriate time to obsolescence of reputation may be shorter or longer.

**6 Conclusions and future research**

On online crowdsourcing labour market platforms, job-seekers' reputation histories remain accessible for unlimited time. In these markets' reputation systems, an experimental lab study examined the presentation of job-seekers' obsolete reputation and how it influences hiring. The findings show that current reputation systems hardly visualize obsolete reputation and scarcely encourage employers to disregard it. As a consequence, employers risk misjudging job-seekers' true current qualification. Job-seekers afflicted by an inadequately presented obsolete reputation get less likely hired resulting in undesirable market outcomes.

Empowering employers to ignore job-seekers' obsolete reputation, online crowdsourcing labour market platforms should graphically visualize time in reputation profiles. Time cues based on the graphical timeline metaphor proofed to make employers better aware of the need to disregard obsolete reputation compared to the temporal order cues implemented on current platforms. The paper contributed by issuing recommendations on improving the display of obsolete reputation in reputation systems and discussing the economic and social benefits of visible time display in online crowdsourcing labour markets. Future research should focus on designing additional visual time cues and develop reusable graphical user interface (GUI) library components. GUI components could simplify the widespread deployment of visual time cues across online contexts presenting people's online reputation. As reputation histories are retained online increasingly longer, future work should devote more attention to the economic and social impacts of obsolete online reputation.

**Figures**

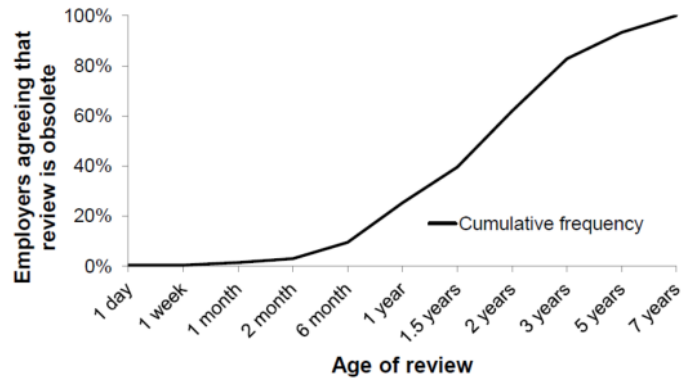

Figure 1. Employer consensus about time to review obsolescence.

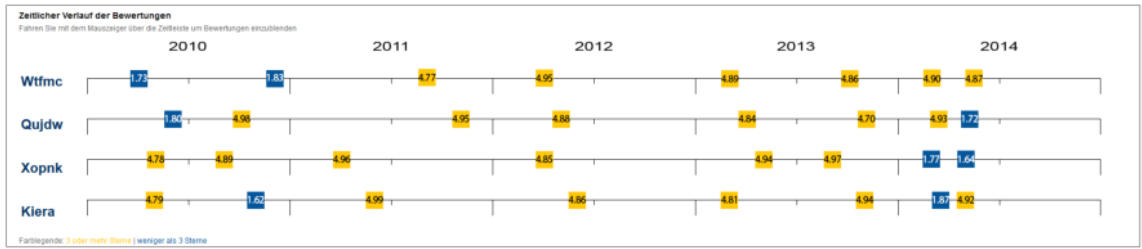

Figure 2. Timelines cue.

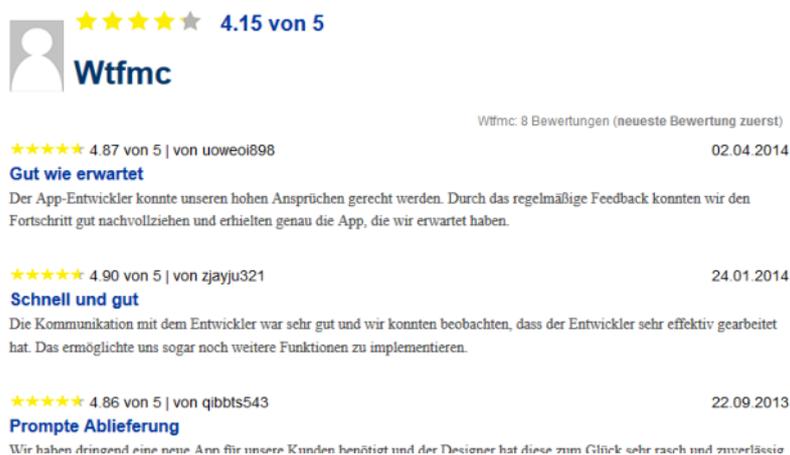

Figure 3. Excerpt of an app designer's reputation profile.

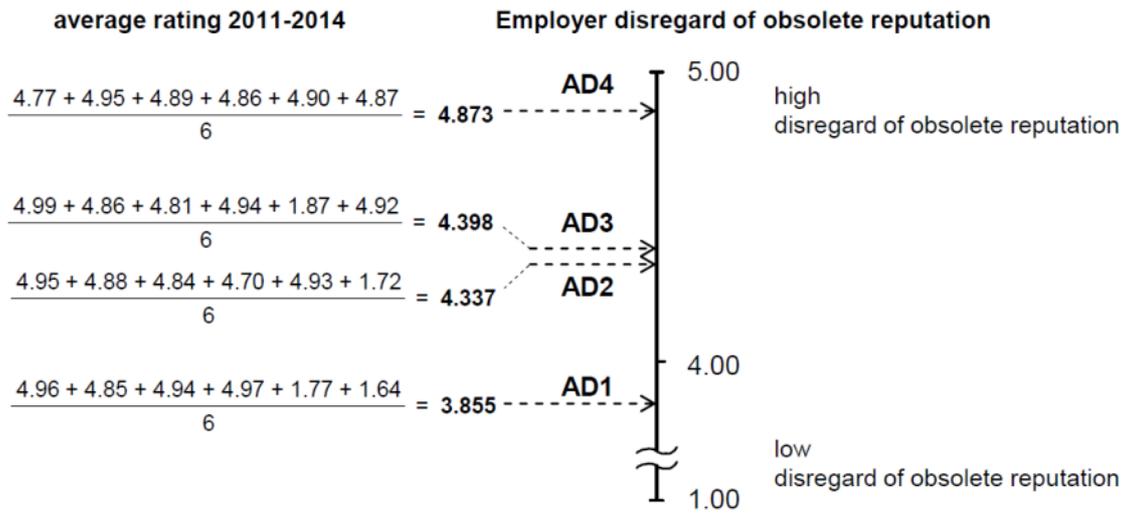

Figure 4. Operationalization of employer disregard of obsolete reputation.

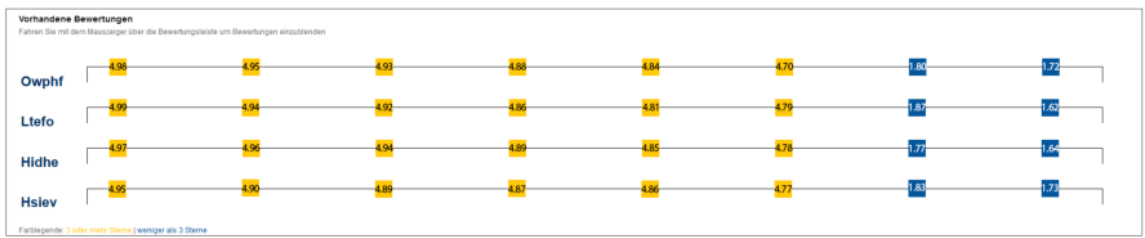

Figure 5. Graphical representations of profiles containing no time information (CC3).

**Tables**

| Year | Reputation profile | | | | | | | | Employer consensus of review obsolescence |
|---|---|---|---|---|---|---|---|---|---|
| | AD1 | | AD2 | | AD3 | | AD4 | | |
| | Date | Rating | Date | Rating | Date | Rating | Date | Rating | |
| 2010 | 04/18 | 4.78 | 05/16 | 1.80 | 04/29 | 4.79 | 03/22 | 1.73 | Obsolete |
| | 08/21 | 4.89 | 09/20 | 4.98 | 10/22 | 1.62 | 11/06 | 1.83 | |
| 2011 | 03/10 | 4.96 | 10/15 | 4.95 | 05/02 | 4.99 | 08/06 | 4.77 | Not obsolete |
| 2012 | 03/18 | 4.85 | 04/19 | 4.88 | 10/22 | 4.86 | 08/08 | 4.95 | |
| 2013 | 04/28 | 4.94 | 03/03 | 4.84 | 02/03 | 4.81 | 02/10 | 4.89 | |
| | 08/18 | 4.97 | 10/27 | 4.70 | 10/07 | 4.94 | 09/22 | 4.86 | |
| 2014 | 02/01 | 1.77 | 02/16 | 4.93 | 03/10 | 1.87 | 01/24 | 4.90 | |
| | 03/11 | 1.64 | 04/21 | 1.72 | 04/12 | 4.92 | 04/02 | 4.87 | |
| Avg. total rating | | 4.10 | | 4.10 | | 4.10 | | 4.10 | All reviews |
| Avg. non-obsolete rating | | 3.855 | | 4.337 | | 4.398 | | 4.873 | Only not obsolete reviews (2011-2014) |

Table 1. Obsolescence of reviews contained in reputation profiles.

| Age (years) | M = 23.31/23.34*, SD = 4.13/4.236*, min = 18/18*, max = 53/53* |
|---|---|
| Sex | female = 56.4%/55.5%*, male = 43.6%/44.5%* |
| Occupation[1] | university student (97.6/97.1%*), high-school student (0.3%/0.3%*), part-time employed (28.7%/28.3%*), full-time employed (2.7%/2.4%*), unemployed (0.6%/0.8%*), other occupation (4.2%/3.9%*) |
| Online market transactions | never (6.0%/5.2%*), less than 1x/year (6.6%/6.8%*), at least 1x/year (49.2%/48.9%*), at least 1x/month (30.4%/30.9%*), at least 1x/week (7.8%/8.2%*) |
| App usage[2] | very often (70.7%/70.7%*), often (18.2%/17.9%*), sometimes (7.0%/6.7%*), rarely (3.8%/3.9%*), never (0.3%/0.8%*) |

[1] multiple selection possible, [2] of those 314/358* participants owning a smartphone or tablet PC, * … asterisked values represent the untrimmed sample (n*=382) before data cleaning.

Table 2. Demographic characteristics of the sample.

|  | Treatment conditions (TC) | |
|---|---|---|
|  | Timelines cue | |
| **Temporal order cue** | no | yes |
| recent review first | **TC1** (n=53, n*=63)<br>M=4.534, SD=0.400<br>M*=4.499, SD*=0.392 | **TC2** (n=54, n*=61)<br>M=4.760, SD=0.237<br>M*=4.724, SD*=0.268 |
| oldest review first | - | **TC3** (n=53, n*=61)<br>M=4.660, SD=0.310<br>M*=4.664, SD*=0.300 |
|  | Control conditions (CC) | |
|  | Other graphical interface representation | |
| **Reviews in other order** | no | yes |
| random order | **CC1** (n=49, n*=65)<br>M=4.404, SD=0.358<br>M*=4.410, SD*=0.346 | - |
| highest rated review first | **CC2** (n=61, n*=63)<br>M=4.392, SD=0.335<br>M*=4.400, SD*=0.335 | **CC3** (n=65, n*=69)<br>M=4.427, SD=0.377<br>M*=4.423, SD*=0.376 |

* … Asterisked values represent the untrimmed sample (n*=382) before data cleaning.

Table 3. Distribution of employer disregard of obsolete reputation across the conditions.

| Step | 1 | 2 | 3 | 4 | 5.a (TC1, 2) | | 5.b (TC1,2,3) | |
|---|---|---|---|---|---|---|---|---|
| Predictor | β, (B), [SE] | β, (B), [SE] | β, (B), [SE] | β, (B), [SE] | β, (B), [SE] | VIF | β, (B), [SE] | VIF |
| Scenario involvement | -0.022 (-0.009) [0.022] | -0.043 (-0.017) [0.021] | -0.068 (-0.026) [0.020] | -0.078 (-0.030) [0.019] | -0.078 (-0.030) [0.019] | 1.074 | -0.079 (-0.031) [0.019] | 1.074 |
| Choice involvement | 0.040 (0.020) [0.028] | -0.006 (-0.003) [0.028] | -0.074 (-0.037) [0.027] | -0.054 (-0.027) [0.026] | -0.054 (-0.027) [0.026] | 1.138 | -0.053 (-0.027) [0.026] | 1.139 |
| Reputation system experience | 0.084 (0.039) [0.026] | 0.071 (0.033) [0.025] | -0.013 (-0.006) [0.025] | -0.008 (0.023) [-0.017] | -0.017 (-0.008) [0.023] | 1.114 | -0.015 (-0.007) [0.023] | 1.117 |
| Forgiveness | | **0.293**\*\*\* (0.106) [0.019] | **0.205**\*\*\* (0.074) [0.019] | **0.172**\*\* (0.062) [0.019] | **0.172**\*\* (0.062) [0.019] | 1.152 | **0.172**\*\* (0.063) [0.019] | 1.152 |
| Rational thinking style | | | **0.323**\*\*\* (0.175) [0.031] | **0.277**\*\*\* (0.150) [0.030] | **0.277**\*\*\* (0.150) [0.030] | 1.352 | **0.276**\*\*\* (0.150) [0.030] | 1.353 |
| Timelines | | | | **0.297**\*\*\* (0.121) [0.020] | **0.277**\*\* (0.113) [0.033] | 2.853 | **0.365**\*\* (0.149) [0.054] | 7.650 |
| Temporal order | | | | | 0.024 (0.010) [0.033] | 2.813 | -0.073 (-0.027) [0.048] | 7.604 |
| Constant | **(4.325)**\*\*\* [0.205] | **(3.950)**\*\*\* [0.209] | **(3.810)**\*\*\* [0.201] | **(3.968)**\*\*\* [0.193] | **(3.970)**\*\*\* [0.193] | | **(3.970)**\*\*\* [0.193] | |
| Corr. R2 | 0.0% | 8.0% | 15.7% | 24.0% | 23.8% | | 23.8% | |
| ΔR2 | 0.4% | 8.0% | 7.7% | 8.3% | -0.2% | | -0.2% | |
| ΔF | 2.306 | **29.908** | **31.058** | **36.800** | 0.089 | | 0.307 | |
| Sig. of step | | \*\*\* | \*\*\* | \*\*\* | | | | |

ß … standardized coefficient, B … estimated coefficient, SE … standard error
*p < 0.05; **p < 0.01, ***p < 0.001, VIF … variance inflation factor
Dependent variable: employer disregard of obsolete reputation, n=335

Table 4. Step-wise linear regression results.

| Hypothesis | Relationship | Empirical support |
|---|---|---|
| H1 | Reputation profile interfaces ordering the reviews by star rating do not influence employer disregard of obsolete reputation. | supported |
| H2 | Reputation profile interfaces ordering the reviews by time increase employer disregard of obsolete reputation. | rejected |
| H3 | Reviews in descending temporal order more strongly encourage employers to disregard obsolete reputation than reviews in ascending temporal order. | rejected |
| H4 | Reputation profile interfaces displaying reviews on graphical timelines increase employer disregard of obsolete reputation. | supported |

Table 5. Summary of empirical support for hypotheses.

# Appendix

| Crowdsourcing labour market platform (URL) | Displays publication date and/or time of review | Sort by time: recent reviews first | Sort by time: oldest reviews first | Sort by rating: highest rating first | Displays graphical time representation |
|---|---|---|---|---|---|
| Elance (www.elance.com) | x | x | | x | |
| Fiverr (www.fiverr.com) | x | x | | x | |
| Freelancer.com (www.freelancer.com) | x | x | x | x | |
| GetACoder (www.getacoder.com) | x | x | | | |
| Guru (www.guru.com) | x | x | | x | |
| iFreelance (www.ifreelance.com) | x | x | | x | |
| Upwork (www.upwork.com) | x | x | | | |
| Peopleperhour (www.peopleperhour.com) | x | x | | | |
| Project4Hire (www.project4hire.com) | x | x | | | |
| Twago (www.twago.de) | x | x | | | |
| **Experimental manipulation** | Textual time cue: all manipulations | Temporal order cue | | Highest rated review first | Timelines cue |

Table 6. Visual time cues on crowdsourcing labour market platforms.

**Rotated component matrix:** Principal component analysis with varimax rotation and Kaiser-normalization.

| Item no. | Item | Component 1 | Component 2 | Component 3 | Component 4 |
|---|---|---|---|---|---|
| **Scenario involvement** | Cronbach's α=0.844, CR=0.860, AVE=0.555 | | | | |
| | Such a CampusApp… | | | | |
| SI1 | would be of no concern - would be of concern to me | 0.798 | 0.066 | 0.135 | 0.032 |
| SI2 | would be useless - would be useful | 0.784 | 0.052 | 0.077 | 0.104 |
| SI3 | would be boring - would be interesting | 0.756 | 0.130 | 0.068 | 0.082 |
| SI4 | would be undesirable - would be desirable | 0.876 | -0.010 | 0.048 | 0.058 |
| SI5 | would not be needed - would be needed | 0.748 | 0.062 | 0.013 | -0.100 |
| **Rational thinking style** | Cronbach's α=0.763, CR=0.765, AVE=0.522 | | | | |
| RT1 | I tackled the choice for an app designer systematically. | 0.027 | 0.777 | 0.149 | 0.087 |
| RT2 | I was very aware of my thinking process. | 0.117 | 0.814 | 0.020 | 0.217 |
| RT3 | I arrived at my decision by carefully assessing the information in front of me. | 0.097 | 0.797 | 0.170 | 0.096 |
| **Choice involvement** | Cronbach's α=0.725, CR=0.734, AVE=0.486 | | | | |
| | The right choice of the app designer is… | | | | |
| CI1 | essential - non-essential | 0.101 | 0.128 | 0.810 | 0.091 |
| CI2 | beneficial - not beneficial | 0.047 | 0.017 | 0.740 | 0.105 |
| CI3 | needed - not needed | 0.115 | 0.202 | 0.808 | -0.020 |
| **Forgiveness** | Cronbach's α=0.719, CR=0.719, AVE=0.468 | | | | |
| F1 | With time I move past negative reviews an app designer received a couple of years ago. | 0.080 | 0.270 | 0.025 | 0.767 |
| F2 | Although an app-designer received negative reviews a couple of years ago, I am eventually able to see him or her as a good designer. | -0.010 | 0.155 | 0.053 | 0.775 |
| F3 | I forgive an app designer the mistakes which were described in a couple of years old reviews. | 0.057 | -0.009 | 0.107 | 0.802 |
| F4 | I do not opt for an app designer who received negative reviews a couple of years ago. | Item dropped | | | |
| F5 | If an app designer received negative reviews a couple of years ago, I continue to appraise him or her badly. | Item dropped | | | |
| Bartlett's test on sphericity (df = 91) p < 0.001***, Kaiser-Meyer-Olkin = 0.786 | | | | | |

Table 7. Scales of control variables and their psychometric characteristics.

**References**

Akerlof, G.A. (1970) 'The market for "lemons": Quality uncertainty and the market mechanism', *The Quarterly Journal of Economics*, Vol. 84 No. 3, pp.488-500.

Aldiri, K., Hobbs, D. and Qahwaji, R. (2008) 'The human face of e-business: Engendering consumer initial trust through the use of images of sales personnel on e-commerce web sites', *International Journal of E-Business Research*, Vol. 4 No. 4, pp.58-78.

Ba, S. and Pavlou, P.A. (2002) 'Evidence of the effect of trust building technology in electronic markets: Price premiums and buyer behaviour', *Management Information Systems Quarterly*, Vol. 26 No. 3, pp.243-268.

BLS (2012) Number of jobs held, labour market activity, and earnings growth among the youngest baby boomers: Results from a longitudinal survey. U.S. Bureau of Labour Statistics, USDL-12-1489.

Cao, Q., Duan, W. and Gan, Q. (2011) 'Exploring determinants of voting for the "helpfulness" of online user reviews: A text mining approach', *Decision Support Systems*, Vol. 50 No. 2, pp.511-521.

Coope, U. (2001) 'Why does aristotle say that there is no time without change?', *Proceedings of the Aristotelian Society*, Vol. 101 No. 3, pp.359-367.

De Hert, P. and Papakonstantinou, V. (2012) 'The proposed data protection regulation replacing directive 95/46/ec: A sound system for the protection of individuals', *Computer Law & Security Review*, Vol. 28 No. 2, pp.130-142.

Depken Ii, C.A. and Gregorius, B. (2010) 'Auction characteristics, seller reputation, and closing prices: Evidence from ebay sales of the iphone', *International Journal of Electronic Business*, Vol. 8 No. 2, pp.170-186.

Dini, F. and Spagnolo, G. (2009) 'Buying reputation on ebay: Do recent changes help?', *International Journal of Electronic Business*, Vol. 7 No. 6, pp.581-598.

Ekstrom, M., Garcia, A.C.B. and Bjornsson, H. (2005) 'Rewarding honest ratings through personalised recommendations in electronic commerce', *International Journal of Electronic Business*, Vol. 3 No. 3, pp.392-410.

Garrett, T.A. (2006) 100 years of bankruptcy: Why more americans than ever are filing. Federal Reserve Bank of St. Louis, https://www.stlouisfed.org/publications/bridges/spring-2006/100-years-of-bankruptcy-why-more-americans-than-ever-are-filing. (Accessed Jul 8, 2015)

Goldberg, L.R. (1990) 'An alternative "description of personality": The big-five factor structure', *Journal of Personality and Social Psychology*, Vol. 59 No. 6, pp.1216.

GoodTechnology (2014) Good technology mobility index report. Report on app, platform and device preferences from the leader in secure mobility. Good Technology, http://media.www1.good.com/documents/rpt-mobility-index-2014.pdf. (Accessed Oct 3, 2014)

Hong, W., Thong, J.Y.L. and Tam, K.Y. (2007) 'How do web users respond to non-banner-ads animation? The effects of task type and user experience', *Journal of the American Society for Information Science and Technology*, Vol. 58 No. 10, pp.1467-1482.

Howe, J. and Robinson, M. (2006) Crowdsourcing: A definition. http://crowdsourcing.typepad.com/cs/2006/06/crowdsourcing_a.html. (Accessed Oct 3, 2014)

Hsieh, Y.-C. and Chen, K.-H. (2011) 'How different information types affect viewer's attention on internet advertising', *Computers in Human Behaviour*, Vol. 27 No. 2, pp.935-945.

Hu, N., Liu, L. and Zhang, J.J. (2008) 'Do online reviews affect product sales? The role of reviewer characteristics and temporal effects', *Information Technology and Management*, Vol. 9 No. 3, pp.201-214.

Jain, K. and Srinivasan, N. (1990) 'An empirical assessment of multiple operationalizations of involvement', *Advances in Consumer Research*, Vol. 17 No. 1, pp.594-602.

Jøsang, A., Ismail, R. and Boyd, C. (2007) 'A survey of trust and reputation systems for online service provision', *Decision Support Systems*, Vol. 43 No. 2, pp.618-644.

Klein, T.J., Lambertz, C., Spagnolo, G. and Stahl, K.O. (2009) 'The actual structure of ebay's feedback mechanism and early evidence on the effects of recent changes', *International Journal of Electronic Business*, Vol. 7 No. 3, pp.301-320.

Lindley, S., Corish, R., Vaara, E.K., Ferreira, P. and Simbelis, V., (2013), 'Changing perspectives of time in HCI' in *CHI '13 Extended Abstracts on Human Factors in Computing Systems (CHI EA '13)*, ACM, Paris, France, pp.3211-3214.

Manosevitch, E., Steinfeld, N. and Lev-On, A. (2014) 'Promoting online deliberation quality: Cognitive cues matter', *Information, Communication & Society*, Vol. 17 No. 10, pp.1177-1195.

Mantelero, A. (2013) 'The eu proposal for a general data protection regulation and the roots of the 'right to be forgotten'', *Computer Law & Security Review*, Vol. 29 No. 3, pp.229-235.

Matzat, U. and Snijders, C. (2012) 'Rebuilding trust in online shops on consumer review sites: Sellers' responses to user-generated complaints', *Journal of Computer-Mediated Communication*, Vol. 18 No. 1, pp.62-79.

Nielsen, J., (1994), 'Enhancing the explanatory power of usability heuristics' in *SIGCHI Conference on Human Factors in Computing Systems*, ACM, Boston, MA, pp.152-158.

Norman, D.A. (1988) *The psychology of everyday things*, Basic Books, New York.

Novak, T.P. and Hoffman, D.L. (2009) 'The fit of thinking style and situation: New measures of situation-specific experiential and rational cognition', *Journal of Consumer Research*, Vol. 36 No. 1, pp.56-72.

Otterbacher, J., (2009), '"Helpfulness" in online communities: A measure of message quality' in *SIGCHI Conference on Human Factors in Computing Systems (CHI '09)*, Boston, MA, pp.955-964.

Plaisant, C., Milash, B., Rose, A., Widoff, S. and Shneiderman, B., (1996), 'Lifelines: Visualizing personal histories' in *SIGCHI Conference on Human Factors in Computing Systems (CHI '96)*, ACM, Vancouver, BC, pp.221-227.

Prosser, W.L. (1960) 'Privacy', *California Law Review*, Vol. 48 No. 3, pp.383-423.

Resnick, P., Kuwabara, K., Zeckhauser, R. and Friedman, E. (2000) 'Reputation systems', *Communications of the ACM*, Vol. 43 No. 12, pp.45-48.

Ricoeur, P. (1988) *Time and narrative. Translated by blamey, k. And pellauer, d.*, University of Chicago Press.

Rosen, J. (2012) 'The right to be forgotten', *Stanford Law Review Online*, Vol. 64 No. 88, pp.88-92.

Santiago, J., Lupáñez, J., Pérez, E. and Funes, M. (2007) 'Time (also) flies from left to right', *Psychonomic Bulletin & Review*, Vol. 14 No. 3, pp.512-516.

Solove, D. (2006) 'A taxonomy of privacy', *University of Pennsylvania Law Review*, Vol. 154 No. 3, pp.477-560.

Spiekermann, S. (2015) *Ethical it innovation - a value based system design approach*, Taylor & Francis, New York.

Thiry, E., Lindley, S., Banks, R. and Regan, T., (2013), 'Authoring personal histories: Exploring the timeline as a framework for meaning making' in *SIGCHI Conference on Human Factors in Computing Systems (CHI '13)*, ACM, Paris, France, pp.1619-1628.

Thompson, L.Y., Snyder, C.R., Hoffman, L., Michael, S.T., Rasmussen, H.N., Billings, L.S., Heinze, L., Neufeld, J.E., Shorey, H.S., Roberts, J.C. and Roberts, D.E. (2005) 'Dispositional forgiveness of self, others, and situations', *Journal of Personality*, Vol. 73 No. 2, pp.313-360.

Turilli, M. and Floridi, L. (2009) 'The ethics of information transparency', *Ethics and Information Technology*, Vol. 11 No. 2, pp.105-112.

USCensus 'Calculating migration expectancy using acs data' [online] https://www.census.gov/hhes/migration/about/cal-mig-exp.html (Accessed Nov 11, 2014).

Wolford, J. 'Ebay seller sues buyer over negative feedback' *WebProNews* [online] http://www.webpronews.com/ebay-seller-sues-buyer-over-negative-feedback-2013-04 (Accessed Feb 25, 2015).

Zacharia, G., Moukas, A. and Maes, P. (2000) 'Collabourative reputation mechanisms for electronic marketplaces', *Decision Support Systems*, Vol. 29 No. 4, pp.371-388.

Zaichkowsky, J.L. (1985) 'Measuring the involvement construct', *Journal of Consumer Research*, Vol. 12 No. 3, pp.341-352.

Zhu, F. and Zhang, X. (2010) 'Impact of online consumer reviews on sales: The moderating role of product and consumer characteristics', *Journal of Marketing*, Vol. 74 No. 2, pp.133-148.

Zittrain, J. (2010) *The future of the internet - and how to stop it*, Yale University Press, New Haven.